\documentclass[aps,noshowpacs,tightenlines,nobalancelastpage,floatfix,onecolumn]{revtex4}
\usepackage{amssymb}
\usepackage{amsmath}
\usepackage{graphicx}
\usepackage{dcolumn}
\usepackage{bm}

\input{psfig.sty}

\begin{document}

\title{Storage and retrieval of single photons transmitted between remote quantum memories}
\date{\today }
\author{T. Chaneli\`{e}re, D. N. Matsukevich, S. D. Jenkins, S.-Y. Lan, T.A.B. Kennedy, and A. Kuzmich}
 \maketitle \affiliation{School of Physics, Georgia Institute of Technology, Atlanta, Georgia 30332-0430}

\textbf{An elementary quantum network operation involves storing a qubit state in an atomic quantum memory node, and then retrieving and
transporting the information through a single photon excitation to a remote quantum memory node for further storage or analysis. Implementations
of quantum network operations are thus conditioned on the ability to realize such matter-to-light and/or light-to-matter quantum state mappings.
Here, we report generation, transmission, storage and retrieval of single quanta using two remote atomic ensembles. A single photon is generated
from a cold atomic ensemble at Site {\it A} via the protocol of Duan, Lukin, Cirac, and Zoller (DLCZ) \cite{duan} and is directed to Site {\it B}
through a 100 meter long optical fiber. The photon is converted into a single collective excitation via the dark-state polariton approach of
Fleischhauer and Lukin \cite{fleischhauer}. After a programmable storage time the atomic excitation is converted back into a single photon. This
is demonstrated experimentally, for a storage time of 500 nanoseconds, by measurement of an anticorrelation parameter $\alpha $. Storage times
exceeding ten microseconds are observed by intensity cross-correlation measurements. The length of the storage period is two orders of magnitude
longer than the time to achieve conversion between photonic and atomic quanta. The controlled transfer of single quanta between remote quantum
memories constitutes an important step towards distributed quantum networks.}

A quantum network, consisting of quantum nodes and interconnecting channels, is an outstanding goal of quantum information science. Such a
network could be used for distributed computing or for the secure sharing of information between spatially remote parties
\cite{ekert,zeilinger,boschi,briegel,knill,duan}. While it is natural that the network's fixed nodes (quantum memory elements) could be
implemented by using matter in the form of individual atoms or atomic ensembles, it is equally natural that light fields be used as carriers of
quantum information (flying qubits) using optical fiber interconnects. The matter-light interface seems inevitable since the local storage
capability of ground state atomic matter cannot be easily recreated with light fields. Interfacing material quanta and single photons is
therefore a basic primitive of a quantum network.

The potential of atomic ensembles to serve as quantum memories has recently attracted considerable attention
\cite{raymer,lukin,fleischhauer,duan,kuzmich3,polzik}, spawning two distinct lines of research. In one, using the physics of ``slow light"
propagation in an optically thick atomic ensemble, weak coherent laser pulses have been stopped and retrieved in a controlled fashion
\cite{fleischhauer,hau0,phillips,hau}. In the other, motivated by the seminal proposal of Duan, Lukin, Cirac, and Zoller (DLCZ) \cite{duan},
correlated pairs of photons and single photons have been produced from an atomic ensemble \cite{kuzmich2,eisaman,jiang,chou,eisaman1,balic}.
Collective atomic qubits, atom-photon entanglement, and quantum state transfer from atomic to photonic qubits have also been demonstrated
\cite{matsukevich}. These initial experimental demonstrations within the DLCZ paradigm were beset by short atomic coherence times, of the order
of the laser pulse length. In contrast, recent advances in atomic ensemble research \cite{matsukevich1} allow for long quantum memory times, in
excess of ten microseconds in the present work, more than two orders of magnitude longer than the duration of the laser pulses involved in the
protocols.

Here we report the synthesis of these two lines of research by demonstrating the generation, transmission, storage and retrieval of single
photons using remote atomic ensembles as quantum memories. The essential ingredient which enables the completion of this synthesis, and which we
report here, is the ability to convert single photons into single collective atomic excitations. In our experiment the remote quantum memories
are based on cold atomic clouds of $^{85}$Rb confined in magneto-optical traps (MOTs) at Sites {\it A} and {\it B}, as shown in Fig.~1. Sites
{\it A} and {\it B} are physically located in adjacent laboratories, with a 100 meter long single-mode optical fiber serving as the quantum
information channel.

\begin{figure}[tb]
\includegraphics[height=5.1cm,width=8.6cm]{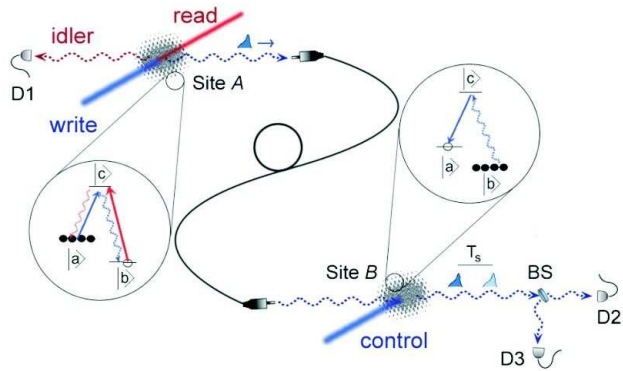}
\caption{{\bf A schematic diagram of our experimental setup demonstrating generation, transmission, storage and retrieval of single photon
excitations of the electromagnetic field}. Two atomic ensembles at Sites {\it A} and {\it B} are connected by a single-mode fiber. The insets
show the structure and the initial populations of atomic levels for the two ensembles. All the light fields responsible for trapping and cooling,
as well as the quadrupole magnetic fields in both MOTs, are shut off during the period of the protocol. The ambient magnetic field at each Site
is compensated by three pairs of Helmholtz coils (not shown). Correlated signal and idler fields are generated at Site {\it A}. The signal field
is transmitted via optical fiber from Site {\it A} to Site {\it B}, where it is converted to atomic excitation, stored for a duration $T_s$, and
subsequently retrieved. A Hanbury Brown-Twiss setup consisting of a beamsplitter BS and two detectors D2 and D3, together with detector D1 for
the idler field, are used to verify the single photon character of the retrieved field.} \label{1}
\end{figure}

Our protocol begins with the generation of single photons at Site {\it A}, using an improved version of the DLCZ approach in the off-axis,
counter-propagating geometry \cite{balic,matsukevich1}. The fiber channel directs the signal field to Site {\it B} where an optically thick
atomic ensemble is prepared in level $|b\rangle $ (right inset in Fig.~1). The signal field propagation in the atomic medium is controlled by an
additional laser field {\it (control)} through the process of electromagnetically-induced transparency (EIT) \cite{harris,scully}. As we deal
with an unpolarized atomic ensemble, we must take into account the Zeeman degeneracy of the atomic levels. Choosing the same circular
polarizations for both the probe and the {\it control} fields allows us to retain transparency, as discussed in more detail in the Supplementary
Information. In Fig.~2 we show the EIT transmission spectrum recorded for a coherent laser probe field instead of the signal field. Evidently, in
the absence of the {\it control} light the probe field is absorbed by the optically thick sample. With the addition of the cw {\it control}
field, the medium is rendered transparent around the $|b\rangle \leftrightarrow |c\rangle $ transition resonance $\Delta =0$.
\begin{figure}[tb]
\includegraphics[width=6.6cm]{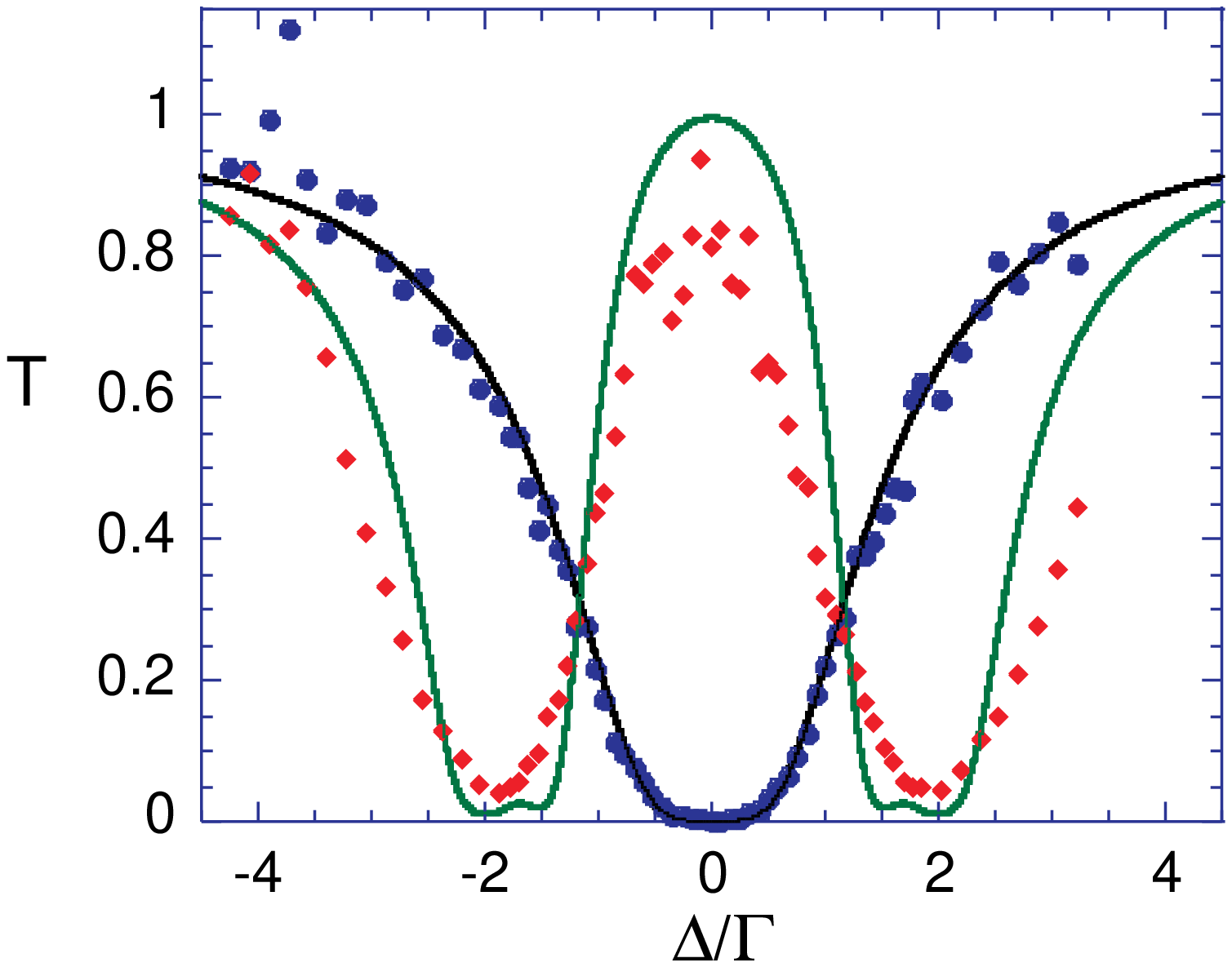}
\caption{{\bf Measured transmission spectra of a coherent probe field as the function of probe detuning in the presence of, and absence of, EIT.}
Data are taken using 700 ns long coherent laser pulses. $T$ is the intensity transmittance, $\Delta $ is the probe detuning and $\Gamma $ is the
decay rate of level $|c\rangle$. In the absence of {\it control} field (circles) the probe is strongly absorbed near resonance, whereas with the
{\it control} field on (diamonds) the medium becomes transparent. Each probe pulse contains on average $0.3$ photons. Each data point is an
average of $2\times 10^5$ experimental trials. The optical thickness $d = 8$ and the {\it control} field Rabi frequency $\Omega = 3 \Gamma $ are
used to obtain the solid curves, based on the theoretical model discussed in the Supplementary Information. } \label{2}
\end{figure}

The {\it control} field strongly modifies the group velocity of the signal field. For a time-dependent {\it control} field, a strong reduction of
the group velocity of the propagating signal field can be understood in terms of a coupled matter-light field excitation known as a ``dark-state
polariton." By adiabatically switching off the {\it control} field, the coupled excitation can be converted into a pure atomic excitation, i.e.,
the signal field is ``stopped" \cite{fleischhauer,phillips,hau}. An important condition to achieve storage is a sufficiently large optical
thickness of the atomic sample, which enables strong spatial compression of the incident signal field \cite{lukin}. In our experiment the
measured optical thickness $d\simeq 8$. Fig.~3 compares our observations with the predictions of a theoretical model. Fig.~3a compares the
propagation of the signal pulse in vacuum and in the atomic medium under conditions of EIT with a cw {\it control} field. The observed pulse
delay under conditions of EIT is about 20 ns, corresponding to more than three orders of magnitude reduction in group velocity. Fig.~3b shows the
effect of turning off the {\it control-storage} field when the signal pulse is approximately centered in the medium, and the subsequent retrieval
of the signal field when the {\it control-retrieval} field is switched back on after a 500 ns storage time. Fig.~3c shows retrieval after a
storage time of 15 $\mu$s. Qualitative agreement of the pulse shapes has been obtained in our theoretical analysis of the protocol using the full
Zeeman structure of the atoms and a classical description of the signal field (Fig.~3d-f).
\begin{figure}[tb]
\includegraphics[width=8.6cm]{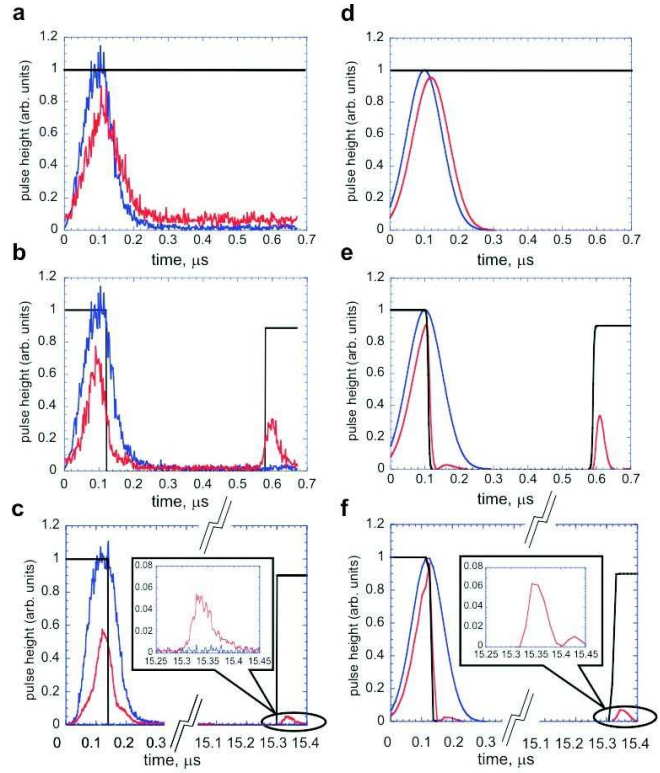}
\caption{{\bf Experimental and theoretical pulse shapes as a function of time, showing EIT, storage and retrieval.} The color code is: {\it
control} field - black, pulse in vacuum - blue, delayed, stored and retrieved field - red. Panel (a) with a cw {\it control} field shows EIT
pulse delay. In panel (b) the {\it control} field is switched off and then on again after 500 ns, shows light storage and retrieval. Panel (c) is
similar to (b) but with a 15 $\mu$s storage. Panels (d), (e), and (f) are corresponding theoretical plots.} \label{3}
\end{figure}

In order to verify the single-photon character of the signal field (a) without storage, and (b) with storage and retrieval, we use a Hanbury
Brown-Twiss detection scheme, employing a beamsplitter followed by two single photon counters, as shown in Fig.~1 \cite{grangier}. To provide
such characterization, we note that classical fields must satisfy a criterion $\alpha \geq 1$ based on the Cauchy-Schwarz inequality
\cite{grangier,uren1}. For an ideally prepared single photon state $\alpha \rightarrow 0$. Here the anticorrelation parameter $\alpha$ is a
function of the storage time $T_s$, and is given by the ratio of various photoelectric detection probabilities which are measured by the set of
detectors D1,D2 and D3  (described in the Methods section):
\begin{equation}
 \alpha (T_s) =\frac{p_1p_{123}}{p_{12}p_{13}}.
\end{equation}

\begin{figure}[tb]
\includegraphics[width=8.0cm]{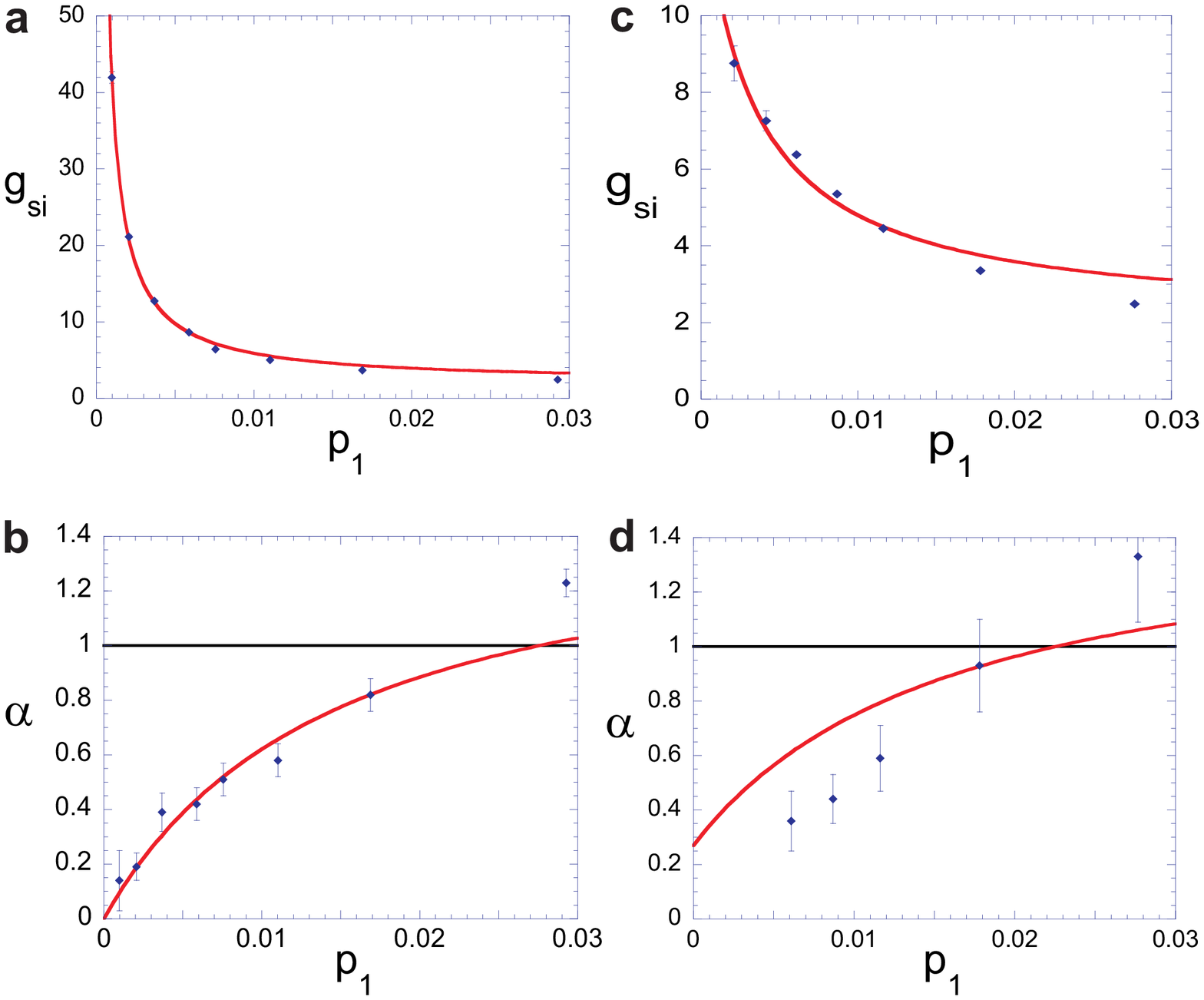}
\caption{{\bf Measured intensity cross-correlation function ${\bf g_{si}}$ and anticorrelation function $ \alpha $ as a function of the idler
photoelectric detection probability ${\bf p_1}$.} Panels (a) and (b) are for the source (propagation in vacuum). Panels (c) and (d) are for
stopped, stored for 500 ns, and retrieved signal field. The solid lines are based on a theoretical model that includes losses and background.
Error bars represent $\pm$ one standard deviation and are based on the statistics of the photoelectric counting events.} \label{4}
\end{figure}

 As an auxiliary measure of signal-idler field
correlations, and as a way to quantify the quantum memory storage time, we also evaluate the normalized intensity cross-correlation function
$g_{si} \equiv (p_{12}+p_{13})/[p_1(p_2+p_3)]$ \cite{clauser,mandel}. In particular, it serves to estimate the total efficiency and background
levels in the experiment, since $g_{si}$ is, by definition, independent of efficiencies whereas $p_1$ is proportional to the overall idler
channel efficiency.

First we measure $g_{si}$ and $\alpha $ without storage at Site {\it B} (i.e., with no atomic sample in place), and the results are displayed in
Fig.~4, a and b, respectively. Next we add an optically thick atomic sample at Site {\it B}, and perform storage of duration $T_s=500$ ns and
subsequent retrieval of the signal field, with results shown in Fig.~4, c and d, respectively. No correction for background or dark counts were
made to any of the experimental counting rates. The curve fits of $g_{si}$ are based on a simple theoretical model, and allow us to obtain the
efficiency in the idler channel and the background contributions to $p_2$ and $p_3$ for the stored signal field. These same values are used to
produce the corresponding theoretical curves in Fig.~4, b and d. The measured values of $\alpha <1$, displayed in Fig.~4, b and d, confirm the
single-photon character of both the source and retrieved signal fields (with the minimum values of $\alpha =0.14 \pm 0.11 $ and $\alpha =0.36 \pm
0.11 $, respectively). Overall, we estimate that the probability $p_s$ for successful generation, transmission, storage, retrieval, and detection
of a signal photon is approximately $p_s \simeq 10^{-5}$ for each trial. The efficiency of photon storage and retrieval $E$ can be estimated as
the ratio of the values of $p_2+p_3$ with and without storage. We find $E \simeq 0.06$, in agreement with the theoretical result shown in Fig.3e.

\begin{figure}[tb]
\includegraphics[width=5.6cm]{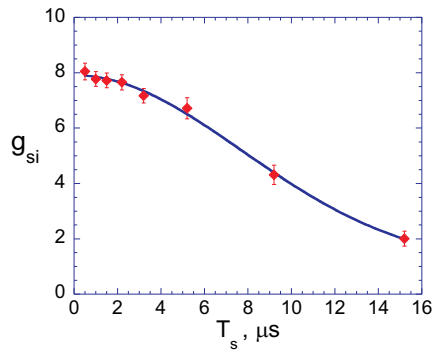}
\caption{{\bf Normalized signal-idler intensity correlation function $g_{si}$ as a function of the storage time $T_s$ at Site {\it B}.} Data
(diamonds) are taken for $p_1=0.0047$, but with a smaller background contribution than that of Fig.4, c and d. The full curve is a fit of the
form $1+ B\exp (-t^2/\tau ^2)$ with $B= 7$ and the collapse time $\tau = 11$ $\mu$s as adjustable parameters. Error bars represent $\pm$ one
standard deviation and are based on the statistics of the photoelectric counting events.} \label{4}
\end{figure}

To investigate the storage capability of our quantum memory at Site {\it B}, we measure $g_{si}$ as a function of the storage time of the signal
field $T_s$  (Fig.~5). A Gaussian fit provides a time constant $\tau = 11$ $\mu$s, which is an estimate of our quantum memory time.  The collapse
is consistent with the Larmor precession of a dark-state polariton in an unpolarized atomic ensemble in a residual magnetic field
\cite{kuzmich2,matsukevich}. Experimentally we attempt to null the uniform, dc component of the magnetic field. A definitive way to distinguish
whether the collapse is due to uniform or non-uniform and ac fields is to measure the damping time of the periodic revivals of the retrieved
signal field at longer storage times. In a uniform magnetic field, undamped revivals of the dark-state polariton should occur at times equal to
$nT_L$, where $T_L$ is the Larmor period for level $|a\rangle $ or $|b\rangle $ and $n$ can be either integer or half-integer, depending on the
direction of the magnetic field relative to the light beam geometry (a synopsis of these ideas is given in the Supplementary Information, with
the full theory presented in Ref.\cite{jenkins}). We have conducted separate experiments with an externally applied magnetic field
\cite{matsukevich2}, which suggest that the collapse in the present experiment is likely due to magnetic field gradients and/or ac fields at the
level of a few tens of mG. However, more extensive investigations to quantitatively determine the temporal and spatial structure of the residual
magnetic field, and the various contributions to it, are ongoing.

We have demonstrated generation, storage and retrieval of single quanta transmitted between two remote atomic ensembles serving as quantum memory
elements. The control of the matter-field interface at the level of single quanta, and at remote sites, is encouraging for further developments
and applications in quantum information science. In particular, the storage of a photonic qubit, with two logical states, would represent a
crucial advance. In order to achieve this, the quantum memory at Site {\it B} would likewise need a second logical state, so as to realize a
collective atomic qubit. Two different approaches for such qubits have already been demonstrated \cite{matsukevich,matsukevich1}. If a second
logical state were added to both quantum memories at Sites {\it A} and {\it B}, generation of remote entanglement of two atomic qubits would be
possible.

{\bf Methods}

{\small To generate single photons at Site {\it A}, we use the DLCZ approach in the off-axis, counter-propagating geometry introduced by Harris
and coworkers \cite{balic}. The insets in Fig.~1 indicate schematically the structure of the three atomic levels involved, $|a\rangle ,|b\rangle
$ and $|c\rangle $, where $\{|a\rangle;|b\rangle \}$ correspond to the $5S_{1/2},F=\{3,2\}$ levels of $^{85}$Rb, and $|c\rangle$ represents the
$\{5P_{1/2},F=3\}$ level associated with the $D_1$ line at 795 nm. The experimental sequence begins with an unpolarized sample of atoms prepared
in level $|a\rangle $ (left inset of Fig.~1). A 160 ns long {\it write} laser pulse tuned to the ${|a\rangle \rightarrow |c\rangle }$ transition
is focused into the MOT with a Gaussian waist of about $400$ $\mu$m. The {\it write} pulse generates a cone of forward Raman-scattered signal
field via the ${|c\rangle \rightarrow |b\rangle }$ transition. We collect a Gaussian mode centered around the momentum $\vec k_s$ that forms an
angle of about $2^{\circ }$ with the {\it write} beam. The {\it write} pulse is so weak that on average less than one photon is scattered into
the collected mode for each pulse. The signal field is coupled into the 100 meter long fiber connecting Sites {\it A} and {\it B}.

For each signal photon emission event, a correlated collective atomic excitation is created in the atomic ensemble. After a delay $\Delta t =200$
ns, a 140 ns long counter-propagating {\it read} laser pulse resonant with the $|b\rangle \rightarrow |c\rangle $ transition illuminates the
atomic ensemble and converts the atomic excitation into the idler field. Under the conditions of collective enhancement, the idler field is
emitted with high probability into the mode determined by the phase-matching condition $\vec k_i = \vec k_w +\vec k_r-\vec k_s$, where $\vec k_i,
\vec k_w$ and $\vec k_r$ are the wave vectors of the idler, {\it write} and {\it read} fields, respectively. The waist of the signal-idler mode
in the MOT is about 150 $\mu$m. The idler field is directed onto a single photon counter D1. Ideally, photoelectric detection of the idler field
projects the quantum state of the signal field into a single photon state. The repetition rate of the experiment is $2 \cdot 10^5$ $s^{-1}$. Each
data point in Fig. 4 involves an average over a time period that varied from several minutes up to 1.5 hours for the data point with the lowest
value of $p_1$ in d.

To measure the photoelectric detection probabilities $p_1,p_2,p_3, p_{13},p_{12}, p_{23}$, and $p_{123}$, the outputs of the detectors are fed to
three ``Stop" inputs of the time-interval analyzer which records the arrival times with a 2 ns time resolution. The electronic pulses from the
detectors D$1$,D$2$,D$3$ are gated for periods $[t_0^i,t_0^i +T_g^i]$, with $T_g^1=140$ ns, $T_g^2=T_g ^3=240$ ns, respectively, centered on the
times determined by the {\it write} and {\it read} (for no storage) or {\it control-retrieval} (for storage) laser pulses. Counts recorded
outside the gating periods are therefore removed from the analysis. The list of recorded events allows us to determine  the single-channel
photoelectric event probabilities $p_i=N_i/M$, where $N_i$ is the total number of counts in the {\it i}-th channel and M is the number of
experimental trials,  (for D$i$, $i=1,2,3$). If photoelectric detections in different channels $i,k,m$ happen within the same gating period, they
contribute to the corresponding joint probabilities $p_{ij}=N_{ij}/M$, where $N_{ij}$ is the total number of coincidences between D$i$ and D$j$,
$i,j=1,2,3$. The joint probability of all three detectors registering a count is given by $p_{123}=N_{123}/M$.}

\section{Appendix}

Here we provide additional information on the theory of electromagnetically induced transparency (EIT) for atoms with Zeeman degeneracy which is
essential for adequate modelling of the system. We also include details of the analysis of the photoelectron counting statistics of the light
fields detected in our experiment, in support of generation, storage, and retrieval of single photon states of the electromagnetic field.

\subsection{Model description}

At Site {\it B} we have an atomic ensemble of $N$ atoms~with three hyperfine levels in a $\Lambda$ configuration labeled by $\left\vert
a\right\rangle $, $\left\vert b\right\rangle $, and $\left\vert c\right\rangle $ with total angular momenta $F_{a}$, $F_{b}$, and $F_{c}$
respectively, as illustrated in Fig. 1. These atoms interact with a classical control field on the $a\rightarrow c$ transition and a signal field
on the $b\rightarrow c$ transition. The dynamics of the interacting system is governed by the Hamiltonian
\begin{equation}
\hat{H}=\hat{H}_{0}+\hat{V}_{AS}+\hat{V}_{AC}+\hat{V}_{AB}%
\end{equation}
where $\hat{H}_{0}=\sum_{\mathbf{k},\lambda}\hbar\omega_{\mathbf{k}}\hat
{a}_{\mathbf{k}\lambda}^{\dag}\hat{a}_{\mathbf{k}\lambda}+\sum_{\mu}\left(
\hbar\omega_{c}\sum_{m_{c}=-F_{c}}^{F_{c}}\hat{\sigma}_{c~m_{c}}^{c~m_{c}%
}\left(  \mu\right)  +\hbar\omega_{a}\sum_{m_{a}=-F_{a}}^{F_{a}}\hat{\sigma }_{a~m_{a}}^{a~m_{a}}\left(  \mu\right)  \right)  $ is the sum of the
free signal field and atomic energies. The slowly varying operators for atom $\mu$, are defined by
$\hat{\sigma}_{s_{2}~m_{2}}^{s_{1}~m_{1}}\left(  \mu\right)
\equiv\left\vert s_{1}m_{1}\right\rangle _{\mu}\left\langle s_{2}%
m_{2}\right\vert \exp\left(  -i\left(  \omega_{s_{2}}-\omega_{s_{1}}\right) \left(  t-z_{\mu}/c\right)  \right)  $. \ Here
$\hat{V}_{AS}=-\sum_{\mu }\mathbf{\hat{D}}_{\mu}\cdot\mathbf{\hat{E}}\left(  \mathbf{r}_{\mu}\right)  $ is the signal field-atom interaction
where $\mathbf{\hat{D}}_{\mu}$ is the dipole operator for atom $\mu$ and $\mathbf{\hat{E}}\left(  \mathbf{r}_{\mu }\right)  $ is the quantized
signal field operator at the atom's position $\mathbf{r}_{\mu}$. The atoms are influenced by the classical control field via the interaction
$\hat{V}_{AC}=-\hbar\Omega\left(  t\right)  \sum
_{m_{a}=-F_{a}}^{F_{a}}C_{m_{a}~r~m_{a}+r}^{F_{a}~1~F_{c}}\sum_{\mu}%
\hat{\sigma}_{a~m_{a}}^{c~m_{a}+r}\left(  \mu\right)  +h.c.$, where $\Omega(t)C_{m_{a}~r~m_{a}+r}^{F_{a}~1~F_{c}}$ is the Rabi frequency of the
control field with polarization $r$. We also account for the presence of a weak, uniform DC magnetic field through the Zeeman interaction
\begin{equation}
\hat{V}_{AB}= \mu_{B}\sum_{\mu}\left(  g_{a}\mathbf{\hat{F}}_{\mu}^{\left(
a\right)  }+g_{b}\mathbf{\hat{F}}_{\mu}^{\left(  b\right)  }+g_{c}%
\mathbf{\hat{F}}_{\mu}^{\left(  c\right)  }\right)  \cdot\mathbf{B}\text{,}%
\end{equation}
where $\mathbf{\hat{F}}_{\mu}^{\left(  s\right)  }\equiv\hat{P}_{s}%
\mathbf{\hat{F}}_{\mu}\hat{P}_{s}$ is the projection of the atomic angular momentum onto the hyperfine level $s$. As we shall demonstrate later,
the associated Larmor precession serves as a possible explanation for the reduction of read-out efficiency of stored photons over long storage
times.

Before the signal field interacts with the ensemble, each atom is prepared in an unpolarized state. We identify this as the mixed atomic
\textquotedblleft vacuum\textquotedblright\ state, $\hat{\rho}_{vac}$, as there are no spin coherence excitations present:
\begin{equation}
\hat{\rho}_{vac}\equiv\frac{1}{\left(  2F_{b}+1\right)  ^{N}}%
{\displaystyle\bigotimes\limits_{\mu=1}^{N}}
\sum_{m_{\mu}=-F_{b}}^{F_{b}}\hat{\sigma}_{b~m_{\mu}}^{b~m_{\mu}}\left(
\mu\right)  \text{.}%
\end{equation}
The signal field is described by the slowly varying photon density operator, $\mathbf{\hat{\Phi}}\left(  z,t\right) =\sum_{\beta }\xi _{\beta
}\hat{\Phi}_{\beta }\left( \mathbf{r},t\right) =\sum_{\beta }\xi _{\beta }\left( i/\sqrt{\mathcal{V}}\right) \sum_{q}\exp\left( i\left(
\mathbf{q\cdot r}-\omega_{c}\right) t\right)  \hat{a}_{\mathbf{q}^{\prime },\beta }$, where $\mathbf{q}^{\prime}=\mathbf{q+}\left(
\omega_{c}/c\right) \mathbf{\hat{z}}$ and $\beta =\pm1$ indicates the polarization $\xi _{\pm1}\equiv\mp\left( 1/\sqrt{2}\right) \left(
\mathbf{\hat{x}\pm }i\mathbf{\hat{y}}\right)  $ of the signal field which propagates in the z-direction. The Rabi frequency on the
$c\leftrightarrow b$ transition is defined by $\hbar g=\left( c\left\Vert \hat{D}\right\Vert b\right)  \left(
\hbar\omega_{c}/2\epsilon_{0}\mathcal{V}\right)  ^{1/2}$\cite{Rose}. \ The decay rate of the optical coherence $\Gamma_{cb}$ is one half of the
spontaneous emission rate of the $D_{1}$ transition, which for $^{85}$Rb is $\Gamma_{c}=2\Gamma_{cb}\approx2\pi\times5.75\operatorname{MHz}$. It
is convenient to define several frequencies resulting from the Zeeman shift of the hyperfine levels in a magnetic field, here chosen for
simplicity to lie in the z-direction:
\begin{align*}
\delta_{c} &  \equiv \beta g_{c}\left(  \frac{\mu_{B}}{\hbar}B_{z}\right)  \\
\Delta_{cb} &  \equiv\left(  g_{c}-g_{b}\right)  \left(  \frac{\mu_{B}}{\hbar
}B_{z}\right)  \\
\delta_{a} &  =\left( \beta-r\right)  g_{a}\left(  \frac{\mu_{B}}{\hbar
}B_{z}\right)  \\
\Delta_{ab} &  =\left(  g_{a}-g_{b}\right)  \left(  \frac{\mu_{B}}{\hbar}%
B_{z}\right)
\end{align*}
where $g_{a}$, $g_{b}$, and $g_{c}$ are the Land\'{e} g-factors for the hyperfine levels $a$, $b$ and $c$, respectively.

\subsection{Field Propagation}

The response of the medium to a time-dependent \textit{control} field enables storage of the signal pulse within the medium. Following the
treatment of Fleischhauer and Lukin \cite{fleischhauer}, but incorporating the Zeeman degeneracy of the atomic levels, and assuming a coherent
signal field, we find that the $\beta$-polarization component of the signal field obeys the propagation equation
\begin{equation}
\frac{\partial}{\partial z}\Phi_{\beta }\left(  \Delta,z\right) =i\frac{\Delta}{c}\int d\Delta^{\prime}\left(  \delta\left(  \Delta
-\Delta^{\prime}\right)  +\frac{1}{2}\frac{\omega_{c}}{\Delta}\chi_{\beta }\left(  \Delta,\Delta^{\prime},z\right)  \right)  \Phi_{\beta }\left(
z,\Delta^{\prime}\right)  ,\label{EqPropagation}%
\end{equation}
where the non-local linear susceptibility is given by
\begin{equation}
\chi_{\beta}\left(  \Delta,\Delta^{\prime}\right)  =-\left(  \frac{c}%
{\omega_{c}}\right)  d_{\beta}^{\prime}\left(  z\right)  \sum_{m=-F_{b}%
}^{F_{b}}~X_{m,\beta }^{2}\frac{\Gamma_{cb}K_{m,\beta}^{-1}\left( \Delta,\Delta^{\prime}\right)  }{\Delta^{\prime}+i\Gamma_{cb}},
\end{equation}
where $d_{\beta}^{\prime}\left(  z\right)  $ is the spatial derivative of the optical thickness $d_{\beta}$, which is defined as the negative
logarithm of the on resonance intensity transmittance of a signal with polarization $\beta$, in the absence of the control field. Explicitly,
$d_{\beta}$ is given by the dimensionless quantity
\[
d_{\beta}\left(  z\right)  =6\pi f_{cb}\left(  \frac{c}{\omega_{c}}\right)
^{2}\int_{0}^{z}dz^{\prime}n\left(  z^{\prime}\right)  \frac{1}{2F_{b}+1}%
\sum_{m}\left\vert C_{m~a~m+a}^{F_{g}~1~F_{e^{\prime}}}\right\vert ^{2}%
\]
where $X_{m,\beta}\equiv C_{m~\beta~m+\beta}^{F_{b}~1~F_{c}}/\sqrt
{\sum_{m_{b}}\left\vert C_{m_{b}~\beta~m_{b}+\beta}^{F_{b}~1~F_{c}%
}\right\vert ^{2}}$, and $f_{cb}=\left\vert \left(  c\left\Vert \hat{D}\right\Vert b\right) \right\vert ^{2}/\left(  \left\vert \left(
c\left\Vert \hat{D}\right\Vert b\right)  \right\vert ^{2}+\left\vert \left(  c\left\Vert \hat{D}\right\Vert a\right)  \right\vert ^{2}\right)  $
is the fraction of atoms in the excited state that spontaneously decays into the ground level $b$, and $n\left( z\right)  $ is the number density
of atoms. Furthermore,
\begin{equation}
K_{m,\beta}=\left(  \delta\left(  \Delta-\Delta^{\prime}\right)  -\frac {1}{\sqrt{2\pi}}\frac{L_{m,\beta}\left(  \Delta,\Delta^{\prime}\right)
}{\left( \Delta+\delta_{c}+m\Delta_{cb}+i\Gamma_{cb}\right)  }\right)
\end{equation}
is a kernel whose inverse satisfies the property $\int d\Delta^{\prime\prime }~K\left(  \Delta,\Delta^{\prime\prime}\right)  K^{-1}\left(  \Delta
^{\prime\prime},\Delta^{\prime}\right)  =\int d\Delta^{\prime\prime}%
~K^{-1}\left(  \Delta,\Delta^{\prime\prime}\right)  K\left(  \Delta ^{\prime\prime},\Delta^{\prime}\right)  =\delta\left(  \Delta-\Delta^{\prime
}\right)  $. Here $\Delta$ is the frequency space Fourier variable, with $\Delta=0$ corresponding to a signal component resonant with the
$c\rightarrow b$ transition. We also have
\begin{equation}
L_{m,\beta}\left(  \Delta,\Delta^{\prime}\right)  =\left\vert C_{m+\beta -r~r~m+\beta}^{F_{a}~1~F_{c}}\right\vert ^{2}\frac{1}{\sqrt{2\pi}}\int
d\Delta^{\prime\prime}\Omega\left(  \Delta-\Delta^{\prime\prime}\right) \frac{1}{\left(  \Delta^{\prime\prime}+\left(
\delta_{a}+m\Delta_{ab}\right) \right)  }~\Omega^{\ast}\left(  \Delta^{\prime}-\Delta^{\prime\prime}\right) .
\end{equation}
With this propagation equation, we are able to numerically calculate the propagation of the signal field leading to the various features shown in
Fig. 3 (we assume equal signal and control polarizations $\beta=r$ in order to have EIT), including group delay, pulse storage and retrieval.

\subsection{Linear susceptibility for constant control field}

In the limit of constant control field, the susceptibility reduces to the form
\begin{equation}
\chi_{\beta}\left(  \Delta,z\right)  \approx\frac{1}{2}\left(  \frac{c}{\omega _{c}}\right)  d_{\beta}^{\prime}\left(  z\right)
\sum_{m}~\frac{\left\vert X_{m,\beta}\right\vert ^{2}\left(  \Delta/\Gamma_{c}\right)  }{\left\vert \Omega/\Gamma_{c}\right\vert ^{2}\left\vert
C_{m+\beta-r~\beta~m+\beta }^{F_{a}~1~F_{c}}\right\vert ^{2}-\left(  \Delta/\Gamma_{c}\right)  \left(
\left(  \Delta/\Gamma_{c}\right)  +i/2\right)  }\label{EIT}.%
\end{equation}
Since the Zeeman shifts are small compared to the spontaneous emission rate, they have been ignored in the above expression. \ Notice that the
medium will only exhibit EIT with a judicious choice of atomic hyperfine levels and field polarizations. \ If one of the Clebsch-Gordan
coefficients multiplying the Rabi frequency $\Omega$ vanishes (i.e. there is an excited state $\left\vert c,m+\beta\right\rangle $ not coupled by
the control field to the corresponding ground state $\left\vert a,m+\beta-r\right\rangle $), then a fraction of the atoms will simply absorb the
signal light as if there is no control field present. We point out that for the level scheme used in the experiment, EIT is achieved only when
the signal and control field polarizations are equal ($\beta=r$). The solid curves in Fig. 2 are based on Eq.(\ref{EIT}).

\subsection{Dark state polariton}

By generalising the analysis of Fleischhauer and Lukin \cite{fleischhauer} to an unpolarized degenerate three level system, one can derive
dark-state polariton operators. These operators, which depend on signal and control field polarizations, exist only under conditions of EIT.
Assuming EIT is possible, for real $\Omega$, the dark-state polariton operator of wavevector $\mathbf{q}$ and polarization $\beta$ is given by
\cite{jenkins}
\begin{equation}
\hat{\Psi}_{\beta}\left(  \mathbf{q},t\right)  =i~~\frac{\Omega\hat
{a}_{\mathbf{q+}\left(  \omega_{c}/c\right)  \mathbf{\hat{z},}\beta}%
+i\sqrt{\frac{N}{2F_b +1}}g^{\ast}\sum_{m}\frac{C_{m~\beta~m+\beta}^{F_{b}~1~F_{c}}%
}{C_{m+\beta-r~r~m+\beta}^{F_{a}~1~F_{c}}}\hat{S}_{a~m+\beta-r}%
^{b~m}\left(  \mathbf{q}\right)  }{\sqrt{\Omega^{2}+\frac{N}{2F_b +1}\left\vert g\right\vert
^{2}\sum_{m}\left\vert \frac{C_{m~\beta~m+\beta}^{F_{b}~1~F_{c}}%
}{C_{m+\beta-r~r~m+\beta}^{F_{a}~1~F_{c}}}\right\vert ^{2}}}%
\end{equation}
where
\begin{equation}
\hat{S}_{a~m+\beta-r}^{b~m}\left(  \mathbf{q}\right)  =\sqrt{\frac{2F_{b}%
+1}{N}}\sum_{\mu}\exp\left(  -i\mathbf{q\cdot r}_{\mu}\right)  \hat{\sigma }_{a~m+a-r}^{b~m}\left(  \mu\right)
\end{equation}
and $\mathbf{r}_{\mu}$ is the position of atom $\mu$. The dark-state polariton operator is a linear combination of electric field and a specific
superposition of hyperfine spin coherences.

\subsection{Larmor collapse}

In the presence of a magnetic field, a stored atomic collective excitation Larmor precesses out of the dark-state polariton mode into orthogonal
collective excitations. \ This causes an apparent decoherence which can be calculated by evaluating the number of dark-state polaritons as a
function of time. Assuming that there are exactly $N_{p}\left(  0\right)  $ dark-state polaritons initially, and for a magnetic field oriented in
the $z$-direction, we find the number of dark state polaritons is given by
\begin{equation}
N_{p}\left(  t\right)  =N_{p}\left(  0\right)  \sum_{m}\sum_{m^{\prime}}%
\frac{\left\vert \frac{C_{m~\beta~m+\beta}^{F_{b}~1~F_{c}}}{C_{m+\beta -r~r~m+\beta}^{F_{a}~1~F_{c}}}\frac{C_{m^{\prime}~\beta~m^{\prime}+\beta
}^{F_{b}~1~F_{c}}}{C_{m^{\prime}+\beta-r~r~m^{\prime}+\beta}^{F_{a}~1~F_{c}%
}}\right\vert ^{2}}{\left(  \sum_{m}\left\vert \frac{C_{m~\beta~m+\beta}%
^{F_{b}~1~F_{c}}}{C_{m+\beta-r~r~m+\beta}^{F_{a}~1~F_{c}}}\right\vert ^{2}\right)  ^{2}}\cos\left(  \left(  m-m^{\prime}\right)  \Delta
_{ab}t\right)  \label{EqNp}%
\end{equation}
where again $\Delta_{ab}=\left(  \frac{\mu_{B}}{\hbar}B_{z}\right)  \left(
g_{a}-g_{b}\right)  $. \ For short storage times, we find%
\[
N_{p}\left(  t\right)  \approx N_{p}\left(  0\right)  \exp\left(  -\frac{1}%
{2}\eta^{2}\left(  \Delta_{ab}t\right)  ^{2}\right)
\]
where
\begin{equation}
\eta^{2}=\sum_{m}\sum_{m^{\prime}}\frac{\left\vert \frac{C_{m~\beta~m+\beta }^{F_{b}~1~F_{c}}}{C_{m+\beta-r~r~m+\beta}^{F_{a}~1~F_{c}}}\frac
{C_{m^{\prime}~\beta~m^{\prime}+\beta}^{F_{b}~1~F_{c}}}{C_{m^{\prime}%
+\beta-r~r~m^{\prime}+\beta}^{F_{a}~1~F_{c}}}\right\vert ^{2}}{\left( \sum_{m}\left\vert \frac{C_{m~\beta~m+\beta}^{F_{b}~1~F_{c}}}{C_{m+\beta
-r~r~m+\beta}^{F_{a}~1~F_{c}}}\right\vert ^{2}\right)  ^{2}}\left(
m-m^{\prime}\right)  ^{2}%
\end{equation}
For the level scheme used in the experiment (with $\beta=r$), $\eta ^{2}\approx2.1$. \ We have used this model to estimate the strength of the
magnetic field (assuming it is oriented along the $z$ direction), given the time constant $\tau$ measured experimentally in Fig. 5. We find the
frequency to be
\begin{equation}
\frac{\Delta_{ab}}{2\pi}\approx14%
\operatorname{kHz}%
\text{.}%
\end{equation}
We have used this value in obtaining the theoretical panels in Fig. 3 and find good agreement with the experimental observations. Clearly from
Eq. (\ref{EqNp}), we can predict the revival of dark-polariton number when
$\Delta_{ab}t$ is a multiple of $2\pi$. The revivals occur at $t_{n}%
^{\parallel}=2\pi n/[\left(  \frac{\mu_{B}}{\hbar}B_{z}\right)  \left\vert g_{a}-g_{b}\right\vert ]\approx\pi/[\left(
\frac{\mu_{B}}{\hbar}B_{z}\right) \left\vert g_{a}\right\vert ]$. By contrast, for a magnetic field perpendicular to the $z$ axis, we find
$t_{n}^{\bot}=2\pi/\left[  \left( \frac{\mu_{B}}{\hbar}B_{z}\right)  \left\vert g_{a}\right\vert \right] =2\pi/\left[  \left(
\frac{\mu_{B}}{\hbar}B_{z}\right)  \left\vert g_{b}\right\vert \right]  $, i.e., $t_{n}^{\bot}=2t_{n}^{\parallel}$.

\subsection{Photoelectric counting statistics}

In order to take into account the possibility of sequences of photon pairs, we use a theoretical model based on parametric down-conversion, in
which the annihilation operators for the idler and signal field are transformed as \cite{mandel}:
\begin{align}
\hat a_{i}^{(out)} = \cosh(\eta)\hat a_{i}^{(in)}+\sinh(\eta)\hat
a_{s}^{\dagger(in)},\nonumber\\
\hat a_{s}^{(out)} = \cosh(\eta)\hat a_{s}^{(in)}+\sinh(\eta)\hat a_{i}^{\dagger(in)}.
\end{align}
Here $\eta$ is the Raman gain at Site \textit{A}. We also wish to assess the overall efficiencies and the background levels in our experiment.
Modelling the background to the signal in terms of a coherent field with average photon number $B_{s}$, we find that
\begin{align}
g_{si}=\frac{(1+2\sinh^{2}(\eta))+B_{s}}{\sinh^{2}(\eta)+B_{s}}. \label{g}%
\end{align}

We also determine the anticorrelation parameter $\alpha$ of Grangier \textit{et al.} \cite{grangier}:
\begin{align}
\alpha=\frac{\sinh^{2}(\eta)(4+6\sinh^{2} (\eta))+4B_{s}(1+2\sinh^{2}(\eta
))}{(1+2\sinh^{2} (\eta) +B_{s})^{2}}. \label{alp}%
\end{align}
The singles count rates at detectors D1, D2, and D3 are given by $R_{1}=\epsilon_{1} \sinh^{2}(\eta)$, $R_{2}=|T|^{2} \epsilon_{2} \sinh
^{2}(\eta)$ and $R_{3}=|R|^{2} \epsilon_{3} \sinh^{2}(\eta)$ (assuming that $R_{i} \ll W$, where $W$ is the repetition rate of the experiment);
$T$ and $R$ are the transmission and reflection coefficients of the beamsplitter BS, shown in Fig.~6.

\begin{figure}[ptb]
\begin{center}
\leavevmode  \psfig{file=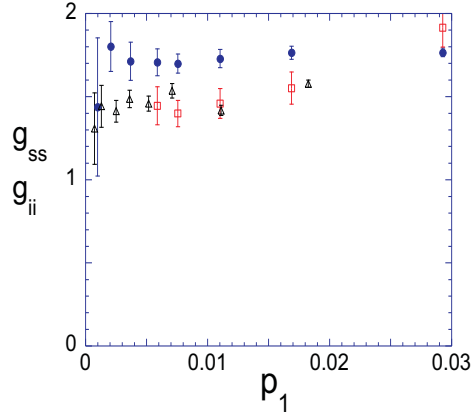,height=2.15in,width=2.4in}
\end{center}
\caption{Normalized intensity autocorrelation functions $g_{ii}$ (triangles) and $g_{ss}$ (circles for the source, squares for the stored and
retrieved field). Uncertainties are based on the statistics of the photon counting
events.}%
\label{TQ}%
\end{figure}

In the absence of the medium we empirically find negligible background $B_{s}%
$. The solid curve in Fig.~4A is based on this model, setting $B_{s}=0$. We find that the best fit to the data in Fig.~4A is given by
$\epsilon_{1} \approx0.039$. The solid line in Fig.~4B is based on Eq.(\ref{alp}) with this value of $\epsilon_{1}$.

For the stored light, we have to account for the fact that a significant fraction of detected signal photons are due to background associated
with the \textit{control-retrieval} pulse. By fitting the data of Fig.~4C to Eq.(\ref{g}), we find $B_{s} \approx0.08$. Substituting this value
into Eq.(\ref{alp}), we obtain the solid curve in Fig.~4D. In order to reduce this background, we have performed initial investigations using an
optically pumped Rb cell to filter out light at the frequency of the control field. In this case we found increased non-classical correlations
between the idler and the stored and retrieved signal photon, e.g., for $T_{s}=500$ ns $g_{si}$ increased from $8 \pm0.2$ to $15.6 \pm1.4$.

In addition, we measure the intensity autocorrelation functions $g_{ss}%
=p_{23}/[p_{2}p_{3}]$ and $g_{ii}$. These are shown in Fig.~1. In order to evaluate the latter, we insert a beamsplitter and additional detector
$D_{a}$ into the path of the idler photon, so that $g_{ii}=p_{1a}/[p_{1}p_{a}]$. Using these together with the measured values of $g_{si}$ shown
in Fig.~4 of
Ref.\cite{chaneliere}, one can evaluate Clauser's parameter $R=g_{si}%
^{2}/[g_{ss}g_{ii}]$. For classical fields $R \leq1$, whereas we observe strong violation of this inequality.

The total measured transmission and detection efficiencies for the idler and signal fields respectively are $w_{i}=0.25 \pm0.03$ and $w_{s}=0.15
\pm0.02$, consisting of the quantum efficiencies of the detectors $0.55 \pm0.05$ and the passive transmission losses accounting for the rest. The
ratio of $h\equiv\epsilon_{i}/w_{i} = 0.16$ indicates the strength of the spatial signal-idler correlations in our source of conditional single
photons at Site \textit{A}, with $h \rightarrow1$ for the ideal case.

{\small \bf Acknowledgements.} {\small This work was supported by NASA, Office of Naval Research Young Investigator Program, National Science
Foundation, Research Corporation, Alfred P. Sloan Foundation, and Cullen-Peck Chair. We thank M. S. Chapman for fruitful discussions and E.T.
Neumann for experimental assistance.}


\end{document}